\documentclass{aa}

\usepackage{graphicx}
\usepackage{subfigure}
\usepackage{txfonts}
\usepackage{longtable}

\newcommand{\sax}{{\it BeppoSAX}}

\newcommand{\xmm}{{\it XMM-Newton}}

\newcommand{\swf}{{\it Swift}}
\newcommand{\sx}{{\it Swift--XRT}}

\begin{document}
\title{X-ray spectral evolution of TeV BL Lac objects:}
  \subtitle{eleven years of observations with \sax, \xmm~and \swf~satellites}

\author{
     F.~Massaro\inst{1}
     \and A.~Tramacere\inst{2}
     \and A.~Cavaliere\inst{1}
     \and M.~Perri\inst{3}
     \and P.~Giommi\inst{3}
}

\institute{
Dipartimento di Fisica, Universit\`a di Roma Tor Vergata
Via della Ricerca scientifica 1 , I-00133 Roma, Italy
\and Dipartimento di Fisica, Universit\`a di Roma La Sapienza,
Piazzale A. Moro 2, I-00185 Roma, Italy
\and ASI Science Data Center, ESRIN, I-00044 Frascati, Italy  
}

\offprints{massaro@roma2.infn.it}
\date{Received ....; accepted ....}

\markboth{F.~Massaro et al.:bla}
{F.~Massaro et al.: bla.}
\abstract{Many of the extragalactic sources detected in $\gamma$ rays at TeV energies are BL Lac objects.
In particular, they belong to the subclass of ``high frequency peaked BL Lacs" (HBLs), as their spectral energy distributions exhibit a first peak in
the X-ray band. At a closer look, their X-ray spectra appear to be generally curved into a log-parabolic shape.
In a previous investigation of Mrk 421, two correlations were found between the spectral parameters.  
One involves the height $S_p$ increasing with the position $E_p$ of the first peak; this was interpreted as a signature of synchrotron emission from relativistic electrons.
The other involves the curvature parameter $b$ decreasing as $E_p$ increases; this points toward statistical/stochastic acceleration processes for the emitting electrons. 
}
{ We analyse X-ray spectra of several TeV HBLs to pinpoint their behaviours in the $E_p-S_p$ and $E_p-b$ planes and to compare them with Mrk 421. 
}
{We perfom X-ray spectral analyses of a sample of 15 BL Lacs.
We report the whole set of observations obtained with the  \sax, \xmm~and \swf~satellites between  29/06/96 and 07/04/07.
We focus on five sources (PKS 0548-322, 1H 1426+418, Mrk 501, 1ES 1959+650, PKS2155-304) whose X-ray observations warrant detailed searching of correlations or trends. 
}
{Within our database, we find that four out of five sources, namely PKS 0548-322, 1H 1426+418, Mrk 501 and 1ES 1959+650, follow similar trends as Mrk 421 
in the $E_p-S_p$ plane, while PKS 2155-304 differs. As for the $E_p-b$ plane, all TeV HBLs follow a similar behaviour. 
}
{The trends exhibited by Mrk 421 appear to be shared by several TeV HBLs, such as to warrant discussing predictions from the X-ray spectral evolution to that of TeV emissions.
}


\authorrunning{F.~Massaro et al.}
\titlerunning{X-ray spectral evolution of TeV BL Lac objects}

\maketitle
\section{Introduction}
Many multiwavelength observations of BL Lac objects support the view that these are active galactic nuclei (AGNs) with relativistc jets pointing close to the line of sight and emitting continuous, Doppler-boosted spectra with limited photon reprocessing at the source (Blandford \& Rees 1978; Urry \& Padovani 1995).

The spectral energy distributions (SEDs) of these sources include two main components: a low-energy component with power peaking in the range from the IR to the X-ray band, and a substantial high-energy component often dominated by $\gamma$ rays.
It is widely agreed that the low-energy component is produced by synchrotron radiation of ultrarelativistic electrons in the jet.
Following the widely entertained synchrotron self-Compton  scenario (SSC; e.g. Jones et al., 1974; Ghisellini \& Maraschi, 1989) the second component may be interpreted as inverse-Compton scattering of the synchrotron photons by the same electron population.

A classification criterion for BL Lacs was suggested by Padovani \& Giommi (1995) on the basis of the location of their first SED peak. This distinguishes the high-frequency peaked BL Lacs (HBLs), for which the peak lies in the UV to the X-ray band, from the low-frequency peaked BL Lacs (LBLs), for which the peak lies in the IR-optical range.

Among the BL Lac objects, several HBLs have been detected at TeV energies, as listed in Table 1.
The synchrotron emissions of such HBLs usually peak in the 0.1-10 keV range, and the extensive X-ray observations now available enable precision studies of their spectral shapes.
In particular, it is widely known (Landau et al. 1986; Fossati et al. 2000; Massaro et al. 2004) that the spectra and therefore the SEDs of BL Lacs often appear to be intrinsically curved.

A recent analysis of Mrk 421 observations performed with  \sax, \xmm~ and ASCA (Tramacere et al. 2007a) has shown two correlations between spectral parameters: the SED peak energy $E_p$ correlates with the peak flux $S_p$ but anticorrelates with the curvature parameter $b$ (as detailed in Sect. 3). These correlations are relevant as signatures of synchrotron emission and  of statistical/stochastic acceleration mechanisms for the emitting electrons, respectively. 

In the present paper, we use a wide set of archival data to investigate the X-ray spectral evolution of other TeV HBLs including all \sax, \xmm~and \swf~published and unpublished archival observations performed between June 1996 and April 2007.
Here, we use the spectra of Mrk 421 (discussed in Tramacere et al., 2007a) as a comparison term for the behaviour of other HBLs.
We do not consider in detail BL Lacertae, recently detected at TeV energies (Albert et al. 2006b), as it belongs to the LBL class; its spectral behaviour will be discussed in a separate paper (Fuhrmann et al., 2007)
\begin{table*}
\caption{A list of the BL Lacs currently detected at TeV energies. Col. (1) reports source names, Cols.(2,3) the right ascension and declination, respectively, Col. (4) gives the redshift (from NED), Cols. (5,6) the galactic column density along the line of sight: ($^{*}$ Lockman \& Savage 1995; $^{**}$ Kalberla et al. 2005), Cols. (7,8,9) report the observing satellite and number of observations, and the final Col. (10) reports the TeV detections. In this paper, we will not pursue Mrk 421 and BL Lac, as anticipated in Sect. 1.}
\begin{flushleft}
\begin{tabular}{|lllccccccc|}
\hline
Name &    $RA$    &   $DEC$     &  $z$  & $N_{H,Gal}^{(*)}$        & $N_{H,Gal}^{(**)}$ &  & Satellite &  & TeV detection\\
            &            &             &       & $[10^{20}cm^{-2}]$ & $[10^{20}cm^{-2}]$&     &      &    &  \\
\hline 
\noalign{\smallskip}
1ES 0229+200& 02 32 48.6 & +20 17 17 & 0.140 & 9.21 &7.69& \sax~(1) &     &     & $HESS^{(1)}$\\     
1ES 0347-121& 03 49 23.2 & -11 59 27 & 0.185 & 3.64 &3.00& \sax~(1) & xmm~(1) & swf~(1) & $HESS^{(1)}$\\    
PKS 0548-322& 05 50 40.6 & -32 16 16 & 0.069 & 2.21 &2.69& \sax~(3) & xmm~(2) & swf~(16) & $HESS^{(1)}$\\     
1ES 1011+496& 10 15 04.1 & +49 26 01 & 0.210 & 0.79 &0.82&        &     & swf~(3) & $MAGIC^{(2)}$\\  
1H 1100-230 & 11 03 37.6 & -23 29 30 & 0.186 & 5.76 &5.60& \sax~(2) & xmm~(2) & swf & $HESS^{(3)}$\\      
\textbf{Mrk 421}&11 04 27.3&+38 12 32 & 0.030 & 1.61 &1.53& \sax & xmm & swf & $Whipple^{(4)}$\\  
Mrk 180     & 11 36 26.4 & +70 09 27 & 0.045 & 1.41 &1.20& \sax~(1) & xmm~(1) & swf~(2) & $MAGIC^{(5)}$\\   
1ES 1218+304& 12 21 21.9 & +30 10 37 & 0.182 & 1.73 &1.81& \sax~(1) & xmm~(1) & swf~(7) & $MAGIC^{(6)}$\\   
1H 1426+428 & 14 28 32.6 & +42 40 21 & 0.129 & 1.38 &1.10& \sax~(1) & xmm~(7) & swf~(7) & $CAT^{(7)}$\\      
1ES 1553+113& 15 55 43.0 & +11 11 24 & ----  & 3.67 &3.72& \sax~(1) & xmm~(1) & swf~(3) & $HESS^{(8)}$\\   
Mrk 501     & 16 53 52.2 & +39 45 37 & 0.033 & 1.71 &1.42& \sax~(11) & xmm~(2) & swf~(10) & $Whipple^{(9)}$\\  
1ES 1959+650& 19 59 59.8 & +65 08 55 & 0.047 & 10.0 &10.1& \sax~(3) & xmm~(3) & swf ~(10)& $Whipple^{(10)}$\\  
PKS 2005-489& 20 09 25.4 & -48 49 54 & 0.071 & 5.08 &3.80& \sax~(2) & xmm~(3) & swf~(3) & $HESS^{(11)}$\\     
PKS 2155-304& 21 58 52.0 & -30 13 32 & 0.116 & 1.69 &1.42& \sax~(3) & xmm~(15) & swf~(22) & $HESS^{(12)}$\\     
\textbf{BL Lac}      & 22 02 43.3 & +42 16 40 & 0.069 & 21.3 &17.1& \sax &     & swf & $MAGIC^{(2)}$\\    
1ES 2344+514& 23 47 04.8 & +51 42 18 & 0.044 & 16.3 &14.2& \sax~(7) &     & swf~(3) & $Whipple^{(13)}$\\    
1H 2356-309 & 23 59 07.9 & -30 37 41 & 0.165 & 1.33 &1.36& \sax~(1) & xmm~(2) &     & $HESS^{(3)}$\\  
\noalign{\smallskip}
\hline
\end{tabular}
\end{flushleft}
(1) http://www.mpi-hd.mpg.de/hfm/HESS/HESS.html; (2) http://wwwmagic.mppmu.mpg.de/index.en.html; (3) Aharonian et al. 2006a; (4) Punch et al. 1992; (5) Albert et al. 2006a; (6) Albert et al. 2006b; (7) Djannati-Ataj et al. 2002; (8) Aharonian et al. 2006b; (9) Quinn et al. 1996; (10) Nishiyama et al. 1999; (11) Aharonian et al. 2005a; (12) Chadwick et al. 1999; (13) Catanese et al. 1998.
\end{table*}
\section{Observations and data reduction}
\subsection{\sax}
Our data set includes the \sax~observations of our sample sources performed
with the narrow field instruments (NFIs): LECS ($0.1 - 10$
keV; Parmar et al. 1997), MECS ($1.3 - 10$ keV; Boella et
al. 1997) and PDS ($13 - 300$ keV; Frontera et al. 1997).
Events for spectral analysis were extracted
following standard procedures. In particular, LECS and MECS events
were selected in circular regions centred at the source position, with radii of 4'
and 8' depending upon the count rate, as indicated by Fiore
et al. (1999). The response matrices and the ancillary response files used in our analysis
have been taken from the BeppoSAX SDC ftp server
(September 1997 release), and background spectra were
taken from the blank field archive.
Standard procedures and selection criteria were applied to the data to avoid the South Atlantic geomagnetic
anomaly, and the solar, bright Earth and particle contaminations
using the SAXDAS (v. 2.0.0) package.

\subsection{\xmm}
Our sources were observed with \xmm~ between May 2000 and May 2006 by means of
all EPIC CCD cameras: the EPIC-PN (Struder et al. 2001), 
and EPIC-MOS  (Turner et al. 2001), operating in different  modes and
with different filters, as described in Appendix A.
Only EPIC-MOS data were reported in this work.

These data are reduced with the same procedure described in Tramacere et al. (2007a). 
Extractions of all light curves, source  and background spectra  are done
using the \xmm~ Science  Analysis System (SAS) v6.5.0.  The calibration
index file  (CIF) and  the summary file  of the observation data file
(ODF) were  generated using updated calibration files (CCF) following the ``User's Guide  to
the \xmm~ Science
Analysis System"  (issue 3.1, Loiseau et  al. 2004) and ``The \xmm~
ABC Guide"  (vers. 2.01, Snowden et al. 2004). 
Event files were produced by the \xmm~ EMCHAIN pipeline.

Light curves  for  every dataset are extracted,  and  all high-background time intervals 
are filtered out to exclude those contaminated  by  solar  flare signals. 
Then, by visual inspection, we select good time intervals far from solar flare peaks and 
with no count rate variations on time scales shorter than 500 seconds.

Photons are extracted from an annular region using
different apertures to minimise  pile-up, which affects MOS data.  The
typical value of the external radius for the annular  region is $40$~$''$.
To filter out pixels affected by significant pile-up, the internal
region was selected by using the EPATPLOT task in {\it XMM-Newton} (SAS) for each observation, 
following the same procedure used in Tramacere et al. (2007a) (for details, see also Loiseau et  al. 2004; Snowden et al. 2004).

In FULL WINDOW images, the background spectrum is extracted from a
circular region with size comparable to the source region, in a
position where sources are not present (typically off axis). 
For other observations, in PARTIAL WINDOW images, no regions are found sufficiently far from
the source for background extraction;
in these cases we use background from blank-field event files (www.sr.bham.ac.uk).
In all cases, we estimate that the average X-ray background flux was
always at $\sim$ 1\% levels of the source flux, resulting in a negligible
contamination of the spectral parameter determination.
A restricted energy range (0.5--10 keV) is used to avoid
possible residual calibration uncertainties.   
To ensure the validity  of  $\chi^2$  statistics,  data are  grouped  by  combining
instrumental channels so that each new bin comprises  40 counts or more, well above the limit for the $\chi^2$ test applicabilty (Kendall \& Stuart, 1979).

\subsection{\swf}
The XRT data analysis is performed with the XRTDAS software (v.~2.1), developed at the ASI Science Data Center (ASDC) and included in the 
HEAsoft package (v.~6.0.2). 
Event files were calibrated and cleaned 
with standard filtering criteria using the \textsc{xrtpipeline} task combined with the latest calibration 
files available in the Swift CALDB distributed by HEASARC. 
Events in the energy range 0.3--10 keV with grades 0--12 (photon counting mode, PC) and 0--2 (windowed timing mode, WT) are used in the analyses 
(see Hill et al. 2004 for a description of readout modes, and Burrows et al. 2005 for a definition of XRT event grades).
To avoid artificially 
high $\chi^2$ values and possible biases in spectral parameter estimations, 
we follow the advice of the \textsc{XRT}  calibration experts, and exclude from 
our analysis the energy channels between 0.4 keV and 0.6 keV (Campana \& 
Cusumano 2006, private communication). 

For the WT mode data, events are selected for temporal and spectral analysis using a 40 pixel wide 
(1 pixel $=2.36$ arcs) rectangular region centred on the source, and aligned along the 
WT one dimensional stream in sky coordinates. 
Background events are extracted from a nearby source-free rectangular region of 40 pixels width and 20 pixels height. 

For PC mode data, when the source count rate  was above $\sim$ 0.5 counts s$^{-1}$ the
data were significantly affected by pile-up in the inner part of the point spread function (PSF). 
To remove the pile-up contamination, we extract only events contained in an annular region centred on the source. The inner radius of the region was determined comparing the observed PSF profiles with the analytical model derived by Moretti et al. (2005), and tipically has a 4 or 5 pixels size, while the outer radius is 20 pixels for each observations.
For \swf~observations in which the source count rate was below the pile-up limit, events are instead extracted using a 20 pixel radius circle. 
The background for PC mode is estimated from a nearby source-free circular region of 20 pixel radius. 

Ancillary response files are generated with the \textsc{xrtmkarf} task 
applying corrections for the PSF losses and CCD defects. 
The latest response matrices (v.~009) available in the Swift CALDB are used, and source spectra are binned 
to ensure a minimum of 30 counts per bin in order to utilize the $\chi^{2}$ minimisation 
fitting technique and ensure the validity  of $\chi^2$ statistics.
\section{Spectral analysis}
We perform our spectral analysis with  the {\sc xspec} software package, version 11.3.2 (Arnaud, 1996).
We describe the X-ray continuum with different spectral models:
an absorbed power-law with column density either free, or fixed at the Galactic value; a power-law with an exponential cutoff; a log-parabolic model (Landau et al. 1986; Massaro et al. 2004). The latter two models are absorbed by a Galactic column density (see Table 1).

The log-parabolic model is tested under the form
\begin{equation}
F(E) = K~E^{-a-b~log(E)}~~~[photons~cm^{-2}~s^{-1}~keV^{-1}],
\end{equation}
or the alternative SED representation
\begin{equation}
S(E) = S_p~10^{-b~log^2(E/E_{p})}~
\end{equation}
with $S_p=E_{p}^{2}\, F(E_p)$. 
After Eq. (2), the values of the parameters $E_p$ (the location of the SED energy peak), $S_p$ (the peak height), and $b$ (the curvature parameter) can be estimated independently in the fitting procedure (Tramacere et al. 2007a).
In all models with fixed Galactic column density, we use $N_H$ values from the LAB survey (Kalberla et al., 2005) or from (Lockman \& Savage 1995), as reported in Table 1. We find consistent spectral parameters (within a 1$\sigma$ interval) from these two column densities.
Tables A.1, A.2 and A.3 report the results from the spectral analyses performed with the standard $N_H$ values from Lockman \& Savage (1995).

Typically, the X-ray spectra of HBLs appear to be featurless and curved (Giommi et al. 2005; Perri et al. 2007) over a broad energy range. 
Absence of spectral features related to any absorbing material was also recently confirmed by Blustin et al. (2004), who performed a detailed analysis of \xmm~RGS spectra of four sources in our sample, namely 1H 1219+301, 1H 1426+428, Mrk 501 and PKS 0548-322.

With an absorbed power-law model we usually obtain unacceptable values of $\chi^2_r$;
even when we leave the low energy absorption as a free parameter, these models are not adequate to describe the high energy end of the X-ray spectra.
Adding to a power-law model, a high energy exponential cutoff $E_c$ corrects the residuals at high energies, however, values of $E_c$ beyond the instrumental energy range are often obtained.
In a few cases where we find an exponential cutoff within the observed energy range, the $\chi^2_r$ values are significantly higher than those found with the log-parabolic model.
In some cases, poor statistics (due to short observational exposures) or restricted instrumental energy range (\xmm~and \swf~relative to \sax), combined with the location of the SED peak outside the observational energy range, make it difficult to evaluate a possible spectral curvature; here, the single power-law model constitutes an acceptable description of  the X-ray spectra.

Analyses of long exposure pointings are performed using time-resolved spectra, as described in Tramacere et al. (2007a), to avoid averaging significant spectral variations while still conserving a sufficient number of counts per observation as to evaluate the spectral curvature.
Results of our spectral analyses are reported in Appendix A; the statistical uncertaintes quoted refer to the 68\% confidence level (one Gaussian standard deviation). 
Observations with less than 30 energy bins after the rebinning procedure or with less than 800 seconds of exposure were not considered for these analyses. 

In conclusion, in agreement with other previous X-ray analyses (Massaro et al. 2004; Massaro et al. 2006; Tramacere et al. 2007a; Tramacere et al. 2007b; Perlman et al. 2005), we find that for about 70\% of all observations the best description for synchrotron spectra of HBLs close to their peak energy is provided by a log-parabolic model. The percentage increases to 99\% for observations with exposures longer than 2000 seconds.
\section{Results}
Two significant correlations between spectral parameters of the log-parabolic model were found by Tramacere et al. (2007a) from studying the X-ray synchrotron emission of Mrk 421.
Specifically, $S_p$ increases with $E_p$ while the curvature $b$ decreases.
Here, we look for any similar correlations or trends in other TeV HBLs.
For effective comparisons it is necessary to make cosmological corrections, even though the redshift range of these sources is rather narrow.
In the log-log representation, the redshift rescaling corresponds to a translation of the spectral energy distribution to higher energies with a increased height.  
The curvature parameter $b$, as defined by Eqs.(1) and (2), is not afffected while the other parameters, $E_p$ and $S_p$, are.

In particular, the rest frame energy peak $E_p^*$ is given by
\begin{equation}
E_p^* = (1+z)~E_p. 
\end{equation}

In addition, noting that the value of $S_p$ is proportional to the bolometric emitted flux, we compare the rest frame powers of BL Lacs in terms of 
the isotropic luminosity peak energy $L_p^*$:
\begin{equation}
L_p^* \simeq 4\pi D_L^2 S_p.
\end{equation}
Here, the luminosity distance $D_L$ of our sources is given by (Peebles, 1993):
\begin{equation}
D_L =  \frac{c}{H_0} (1+z) \int_{0}^{z}\frac{dz}{\sqrt{\Omega_M (1+z)^3+\Omega_{\Lambda}}},
\end{equation}
using a flat cosmology with $H_0=72$ km/(s Mpc), $\Omega_{M}=0.27$ and
$\Omega_{\Lambda}=0.73$ (see Spergel et al., 2007).

To search for trends, one needs at least 10 observations with $E_p$, $S_p$, $b$ well estimated, a requirement satisfied by only 5 of our sources, namely PKS 0548-322, 1H 1426+418, MRK 501, 1ES 1959+650 and PKS 2155-304.
For these sources we perform spectral analyses to evaluate independently $E_p$, $S_p$, $b$, to which we apply the redshift corrections discussed above.

We investigate the presence of trends by evaluating the linear correlation coefficient $r_{log}$ between the logarithms of spectral parameters.
For Mrk 421, the results are $r_{log}=0.67$ and $r_{log}=-0.67$ for the $E_p^*-L_p^*$ and $E_p^*-b$ relations, respectively.
In Figs. 1, 2, 4, 6 and 8, we plot the values of the spectral parameters, with their uncertanties, for each of the five sources and include those of Mrk 421 for comparison.

Figure 1 shows the results for PKS 0548-322; it is worth noting that in this source both $E_p^*$ and $L_p^*$ vary in a narrower range compared to Mrk 421.
PKS 0548-322 follows the same trend of Mrk 421 on both the $E_p^*-L_p^*$ and $E_p^*-b$ planes, with correlation coefficients $r_{log}=0.61$ and $r_{log}=-0.60$, respectively.
\begin{figure}[!htp]
\includegraphics[height=9.5cm,width=9.5cm,angle=0]{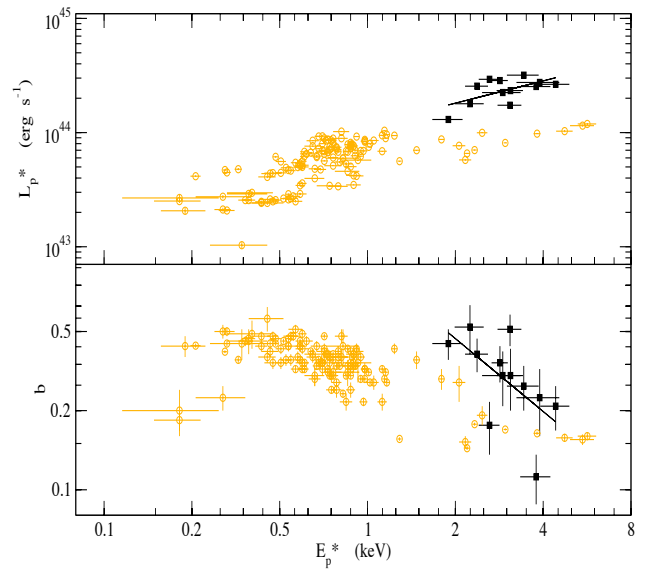}
\caption{$E_p^*-L_p^*$ and $E_p^*-b$ plots for PKS 0548-322 (black filled squares) compared with those of Mrk 421 (orange circles). 
Black lines indicate the regressions underlying the $r_{log}$ correlation coefficient.}
\end{figure}

The source 1H 1426+428 has a similar behaviour to Mrk 421 in the $E_p^*-L_p^*$ and $E_p^*-b$ plots (see Fig. 2), even though it is an order of magnitude brighter than the latter.
In this figure, we also show the \xmm~observation performed on 16 June 2001 (circled in the figure), in which the log-parabolic bestfit indicates a value of $E_p^*$ beyond the instrumental energy range; 
this circumstance makes the formal uncertainty unreliable, and motivates us to exclude this pointing from our statistical analysis.
The observation of \xmm~ on 16 June 2001 appears to confirm the statistical trend in the $E_p^*-b$ plane, but in the $E_p^*-L_p^*$ plane it lies in a different position relative to other pointings. 
As shown in Fig. 3, during this particular pointing, 1H 1426+428 shifted its SED peak energy without large variation of $L_p^*$,  at variance with the following ones by \xmm. 
We find correlation coefficients $r_{log}=0.72$ for the $E_p^*-L_p^*$ relation, and $r_{log}=-0.47$ for the $E_p^*-b$ one; these confirm the similarity to Mrk 421 and to PKS 0548-322.
Note that 1H 1426+428 also covers similar regions on the  $E_p^*-L_p^*$ and $E_p^*-b$ to Mrk 421. 
\begin{figure}[!htp]
\includegraphics[height=9.5cm,width=9.5cm,angle=0]{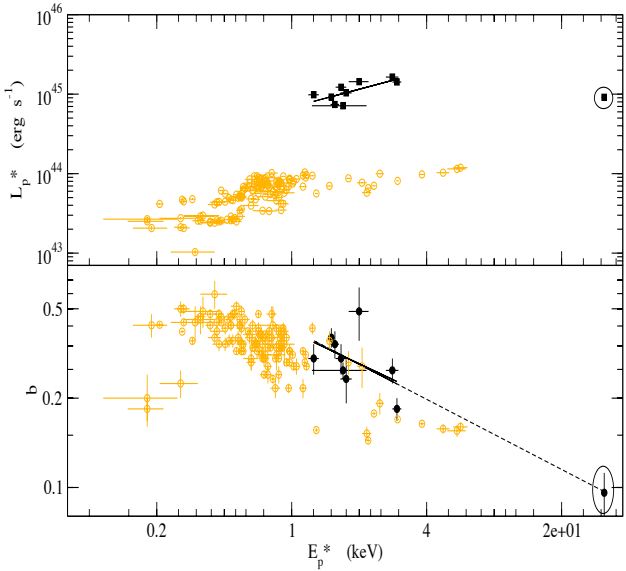}
\caption{$E_p^*-L_p^*$ and $E_p^*-b$ plots for 1H 1426+428 (black filled squares) compared with those of Mrk 421 (orange circles). 
Black lines indicate the regressions underlying the $r_{log}$ correlation coefficient. Circled values refers to the peculiar observation performed on the 16 June 2001 by \xmm~ (see Sect. 4 for details).}
\end{figure}
\begin{figure}[!htp]
\includegraphics[height=9.5cm,width=9.5cm,angle=0]{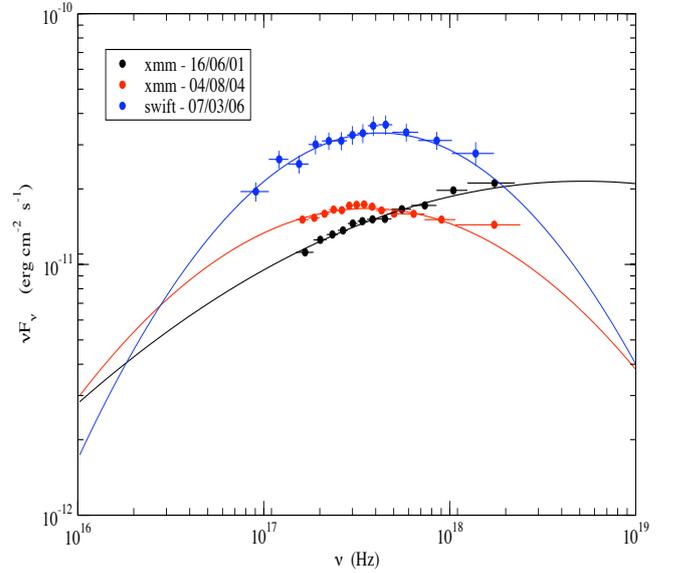}
\caption{The SEDs for three observations of 1H 1426+428 performed by \xmm~and \swf.}
\end{figure}

For Mrk 501 the $E_p^*-L_p^*$ and $E_p^*-b$ plots are shown in Fig. 4. 
Here the range of $E_p^*$ is wider and the luminosities are higher compared to Mrk 421.
Figure 5 shows in detail the strong variability of this source.
The source has similar trends to Mrk 421, with higher correlation coefficients for the $E_p^*-L_p^*$ and $E_p^*-b$ relations, namely,  $r_{log}=0.89$ and  $r_{log}=-0.79$, respectively. 
Figure 5 shows the SEDs relative to three observations performed with all three satellites to show in detail the variations of $E_p$, $s_p$, and curvature $b$.
\begin{figure}[!htp]
\includegraphics[height=9.5cm,width=9.5cm,angle=0]{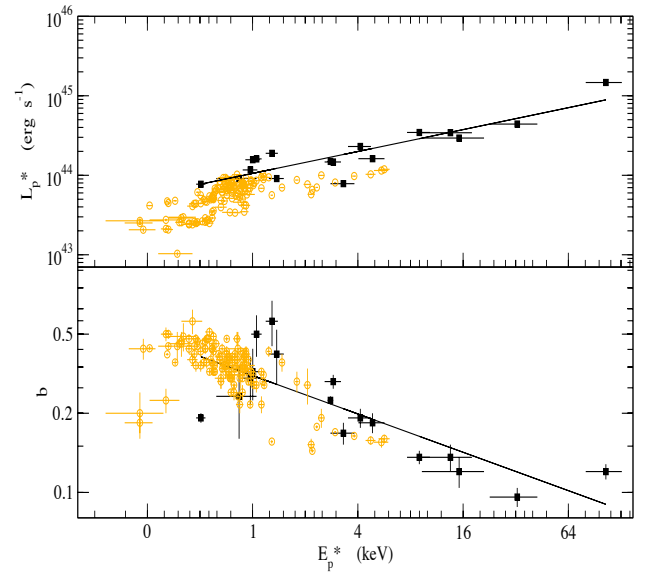}
\caption{$E_p^*-L_p^*$ and $E_p^*-b$ plots for Mrk 501 (black filled squares) compared with those of Mrk 421 (orange circles). 
Black lines indicate the regressions underlying the $r_{log}$ correlation coefficient.}
\end{figure}
\begin{figure}[!htp]
\includegraphics[height=9.5cm,width=9.5cm,angle=0]{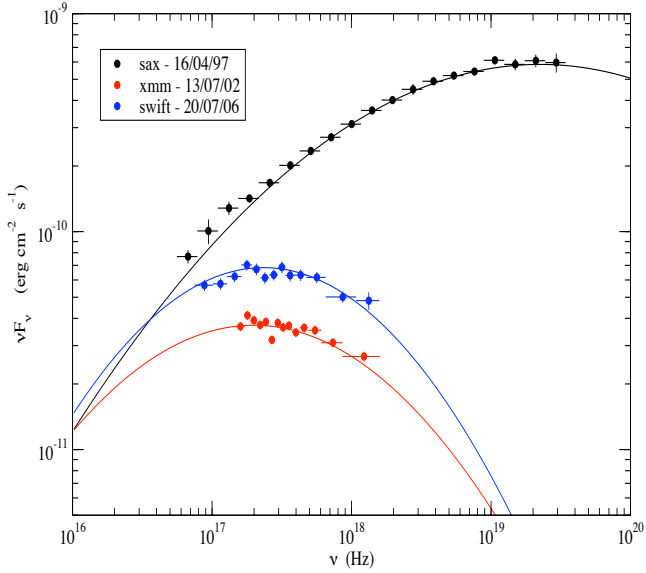}
\caption{The SEDs for three observations of Mrk 501 performed by \sax, \xmm~and \swf.}
\end{figure}

The observations of 1ES 1959+650 cover a narrower subregion of both the $E_p^*-L_p^*$ and the $E_p^*-b$ plane relative to Mrk 421, as shown by Fig 6. 
These observations were mostly performed within ten days during 2006. 
The observation performed on 29 May 2006 (circled) is peculiar as it yields a very high curvature value. This pointing took place at the end of a set of 6 observations, in which
the flux was decreasing; this may represent a phase dominated by cooling, when the estimated value of the curvature could well be affected by an exponential cutoff close to the observed energy range.
\begin{figure}[!htp]
\includegraphics[height=9.5cm,width=9.5cm,angle=0]{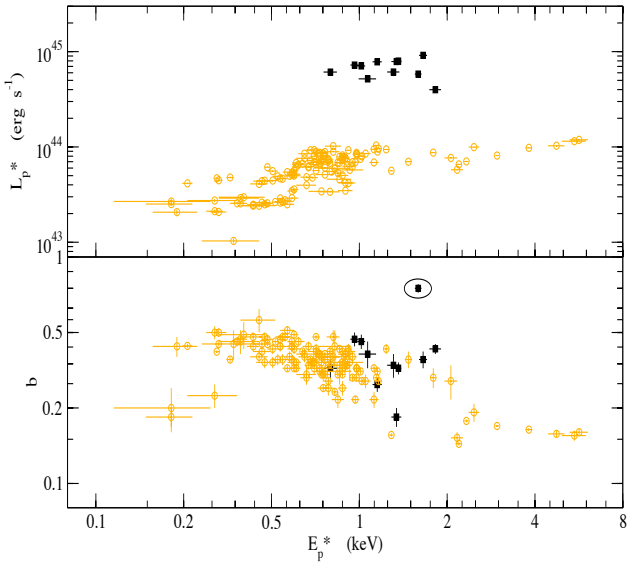}
\caption{$E_p^*-L_p^*$ and $E_p^*-b$ plots for 1ES 1959+650 (black filled squares) compared with those of Mrk 421 (orange circles). Circled values refers to the peculiar observation performed on the 29 May 2006 by \swf~ (see Sect. 4 for details).}
\end{figure}

The source PKS 2155-304 is the truly variant member of our set in a number of respects. 
In fact, the spectral analysis yields a log-parabolic index $a>2$, and relatedly, $E_p$ is less than 1 keV.
It was difficult to evaluate the SED peak location with \sax, \xmm~and \swf~ because it often falls below the observational X-ray range, as shown in Fig. 7.
Such spectra indicate that the X rays constitute the upper end of a synchrotron emission. On the other hand, we never observed a high energy exponential cutoff in our analysis, which confirmes our modelling in terms of a spectral curvature $b$.
The source PKS 2155-304 covers a region in the $E_p^*-b$ plane overlapping that of Mrk 421 in Fig. 8.
On the other hand, the same figure shows that the source does not appear to follow a similar trend in the $E_p^*-L_p^*$ plane.  
A possible explanation for this is
that our X-ray observations may be biased in that we observe the source only with $E_p$ values in the X-rays band, 
corresponding to higher states relative to its average (Tramacere et al., 2007b). 
\begin{figure}[!htp]
\includegraphics[height=9.5cm,width=9.5cm,angle=0]{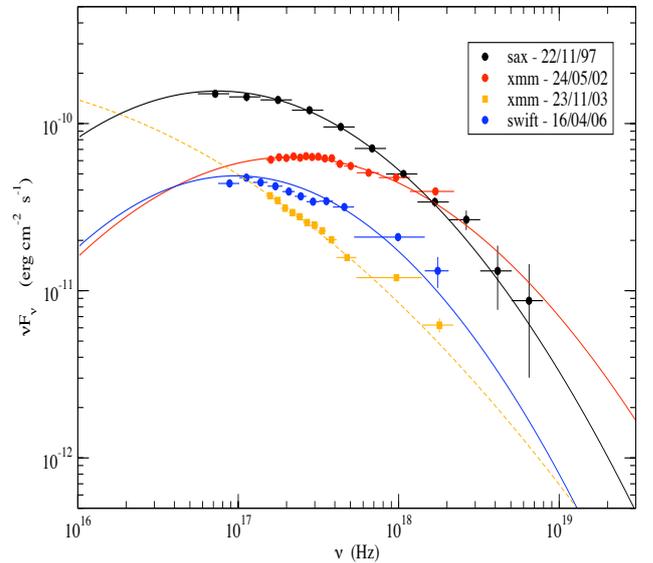}
\caption{The SEDs for four observations of PKS 2155-304 performed by \sax, \xmm~and \swf.}
\end{figure}
\begin{figure}[!htp]
\includegraphics[height=9.5cm,width=9.5cm,angle=0]{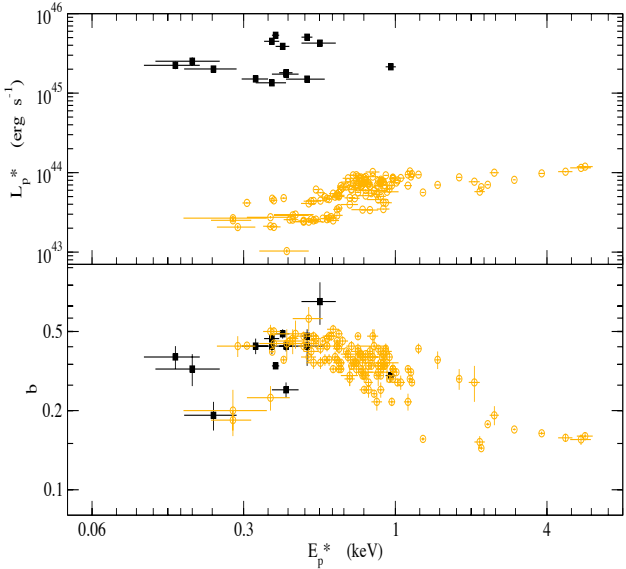}
\caption{$E_p^*-L_p^*$ and $E_p^*-b$ plots for PKS 2155-304 (black filled squares) compared with those of Mrk 421 (orange circles).}
\end{figure}
\section{Discussion}
Correlations between $L_p^*$ and $E_p^*$ provide interesting information concerning
the driver of the source spectral evolution. For example, using a wide dataset of X-ray observations of Mrk 421 we have investigated the effects of varing physical parameters in 
the synchrotron emission, where the dependence of $L_p^*$ on $E_p^*$ may be represented in the form of a
power-law, that is, $L_p^* \propto E_p^{*~\alpha}$  (Tramacere et al., 2007a, and references therein).

In fact, the synchrotron peak is expected to scale as $L_p^* \propto N~\gamma^{2}~B^2~\delta^4$
while the peak energy scales as $E_p^* \propto \gamma^2B~\delta$, in terms of the number $N$ of emitting particles, the
magnetic field $B$,  of the typical electron energy $\gamma mc^2$, and of the beaming factor $\delta$.
Thus, $\alpha=1$ applies\footnote{We take the opportunity to correct here an error in Tramacere et al. (2007a).} when the spectral changes are dominated by variations of
the electron average energy,
$\alpha=2$ applies for  changes of the magnetic field,
$\alpha=4$  if  changes in the beaming factor dominate and
formally, $\alpha=\infty$ (i.e., a vertical line in the $E_p^*-L_p^*$ plane) applies for changes only in the number of emitting particles.

Here, we have presented accurate analyses of the X-ray spectra of several TeV HBLs observed over a period 11 years.
We confirm that these spectra are best described with a log-parabolic model, even though in some cases an acceptable 
fit is also provided by a power-law spectral model absorbed by a Galactic column density.

From our analyses we have derived values of spectral parameters, $E_p$, $S_p$ and curvature $b$, independently.
With the cosmological transformations given by Eqs. (3) and (5), we searched for possible 
correlations, or at least trends, among the spectral parameters.
Five sources (PKS 0548-322, 1H 1426+428, Mrk 501, 1ES 1959+650, PKS2155-304) have enough data to warrant investigating 
in some detail the $E_p^*-L_p^*$ and $E_p^*-b$ relations and comparing them with those found for Mrk 421.

On the other hand, the number of observations for each source in our sample does not allow statistical analyses as detailed as in the case of Mrk 421 (Tramacere et al. 2007a). 
Therefore for these sources it is not yet possible to determine the value of the synchrotron exponent $\alpha$.
Accordingly, we have evaluated only the logarithmic correlation coefficients $r_{log}$ between  $E_p^*-L_p^*$ and $E_p^*-b$ for each source. 

Comparing these values with those evaluated for Mrk 421 we have found that at least three sources (namely PKS 0548-322, 1H 1426+428 and Mrk 501) follow the same trends as Mrk 421 in the  $E_p^*-L_p^*$  plane.
In the case of 1ES 1959+650, our observed spectral parameters cover a smaller region compared to Mrk 421; nevertheless, the trend so outlined is consistent with that of the latter.
Finally, we have found that PKS 2155-304 has again a similar behaviour in the $E_p^*-b$ plane but definitely a different one in the $E_p^*-L_p^*$ plane.

An overall comparison of these similarities is given in Fig. 9 (upper panel). This portrays the $E_p^*-b$ plane for these five sources plus Mrk 421, to show that 
the curvature ranges from about 0.12 to about 0.55 (with the exception of only one pointing of 1ES 1959+650, as discussed above); the correlation coefficient for the sample constituted by these sources is $r_{log}=-0.66$.
Examination of  Tables A1, A2 and A3 indicates that the remaining sources in our sample are consistent with the trend estabilished for Mrk 421 and confirmed by the five HBLs discussed above (see also Fig. 9 lower panel).

Next, we point out two cautionary remarks on biases that may arise when comparing analyses of different sources. 
First, we note the role of the beaming factor. Although Tramacere et al. (2007a) show that for Mrk 421 the beaming factor is unlikely  to be the main driver of the  $E_p^*-L_p^*$ relation, it  may play a subtler role when comparing several sources. 
In fact, both $E_p^*$ and $L_p^*$ depend on $\delta$; this implies that even though for a single source  $\delta$ does not have a large variation, its value may vary significantly from source to source, affecting the $E_p^*-L_p^*$ plot. The same holds for the magnetic field intensity.
A second effect may be given by a poor temporal sampling. 
Sources observed sporadically, with observations covering short temporal intervals, may be representative only of flaring or of low emission states, thus producing an uneven coverage of the parameter space.

Finally, we outline a link between the synchrotron peak and the TeV emissions.
In fact, within a single zone SSC scenario, we expect that synchrotron signatures derived from X-rays observations have counterparts in the TeV energy range, where the inverse Compton peak lies.
In closer detail, we expect the inverse Compton peak height $C_p^*$ and its location $\epsilon_p^*$ in energy to be given by

\begin{equation}
C_p^* \propto N^2 R^{-2} \gamma^4 B^2 \delta^4 \\
\epsilon_p^* \propto  \gamma^4 B \delta~~~~~
(Thomson~regime)
\end{equation}

\begin{equation}
C_p^* \propto N R^{-2} B \delta^4\\
\epsilon_p^* \propto \delta \gamma~~~~~
(Klein-Nishina~regime),
\end{equation}
where $R$ is the size of the emitting region.
It transpires that a definite link exists between the correlations for the variations of the synchrotron peak and variability of the inverse Compton peak,
an issue that we plan to discuss in full in a forthcoming paper.
\begin{figure}[!htp]
\includegraphics[height=9.5cm,width=9.5cm,angle=0]{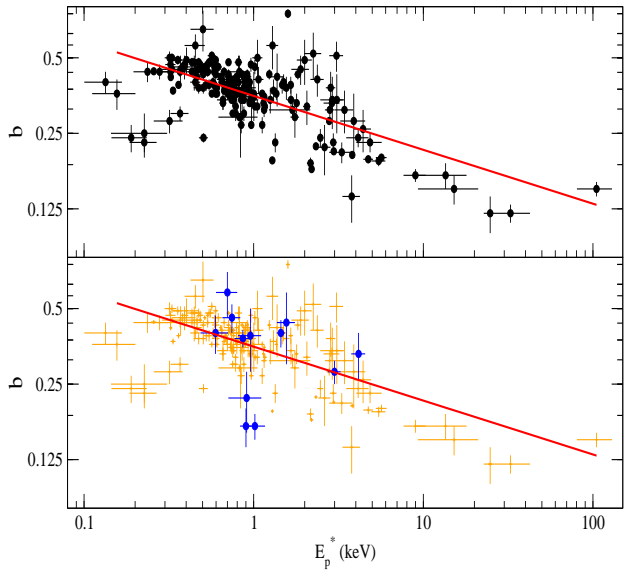}
\caption{Upper panel: the $E_p^*-b$ plot for Mrk 421 and for the five sources analysed in detail in Sect. 5. 
Lower panel: blue points represents the other TeV HBLs with insufficient data to perform a detailed analysis. The above sources are replotted with  orange crosses.}
\end{figure}
\section{Conclusions}
In the present paper, we have conducted an X-ray study of 15 HBL sources, also detected at TeV energies, to investigate their synchrotron emission. Our main results are summarised as follows.

We confirm that the log-parabolic model provides the best description for the X-ray spectra of HBLs, when the statistics are sufficient to warrant evaluating a spectral curvature.
We show that several HBLs detected at TeV energies share the same trends found for Mrk 421 in the $E_p^*-L_p^*$ and $E_p^*-b$ planes.
In particular, for the $E_p^*-b$ relation, all HBLs in our sample cover the same region of these parameter's plane, with the curvature $b$ always decreasing with increasing $E^*_p$.
Finally, we outline a link of the TeV emission with the synchrotron peak in a single zone SSC scenario.
\appendix 
\section{Spectral analysis of the HBL sample}
Tables A.1, A.2 and A.3 report the log of our observations and the values of the spectral parameters we have derived for TeV HBLs in our sample. 

In Table A.1, $LECS$, $MECS$ and $PDS$ columns indicate the exposure time in seconds.
The BeppoSAX spectral analysis of 1ES 0347-121, 1ES 1011+492, PKS 2344-514 is reported in Giommi et al. (2005, also available at http://www.asdc.asi.it/sedentary/).

In Table A.2, $Frame$ indicates the EPIC camera used (M1$=$MOS1 and M2$=$MOS2), the modes (PW$=$partial window and FW$=$full window) and the filter (Th$=$thin, Md$=$medium, Tk$=$thick) used for each pointing (see Sect. 2.2 for details), and the exposure is reported in seconds in the column $Exps$. In Table A.2 capital letters near the observation date indicate a different pointing in the same observation, while lower case letters refer to time resolved spectra (see also Tramacere et al. 2007a). 
The capital letter $F$ in the last column \xmm~table (A.2) indicates that the observation is too contaminated by solar flares to be used in our spectral analysis.

In \swf~Table A.3 the column $Frame$ reports on the observation modality (PC for photon counting and WT for windowed timed, see also Sect. 2.3 for details), and $Exps$ means the exposure time in seconds.

All other columns in each table refer to the log-parabolic model bestfit. 
When the value estimated for a spectral parameter is consistent with zero in a $3\sigma$ interval, the values reported in each table refer to the power-law model bestfit (see Sects. 3 and 4). In these cases, the curvature parameter $b$, the SED peak energy $E_p$ and the corresponding SED peak height $S_p$ cannot be reliably evaluated, and are marked with a dashed line.
Values of $E_p$ are reported in $keV$, the normalisation $K$ in units of $10^{-4} photons~cm^{-2}~s^{-1}~keV^{-1}$ and $S_p$ in units of  $10^{-13} erg~cm^{-2}~s^{-1}$ with $F_X$ denoting the $2-10$ $keV$ flux measured in units of  $10^{-11} erg~cm^{-2}~s^{-1}$.
For spectra with less than 30 bins we report only the estimate of the X-ray flux $F_X$ with a power-law model absorbed by a galactic column density (see Sect. 3 for details).
\begin{acknowledgements}
     We  thank G. Cusumano  for his help in the use of the \sx~ data reduction procedure.
     F. Massaro acknowledges 
     support  by a fellowship of the Italian Space Agency (ASI), 
     in the context of the AGILE Space Mission.  
     We thank our referee for several
     comments helpful toward improving our presentation.
\end{acknowledgements}


\begin{table*}
\caption{\sax~spectral analysis results.}
\begin{tabular}{|llllrrrrrrrr|}
\hline
\noalign{\smallskip}
Obs ID&$Date$&$LECS$&$MECS$&$PDS$&$a$&$b$&$E_p$&$K$&$S_p$&$F_X$&$\chi^2_{r}$\\
\hline  
\noalign{\smallskip}
\textbf{1ES 0229+200}&   &     &     &     &          &          &          &       &            &     &         \\
51472001 & 16/07/01  &17236&68200&33113&1.60(0.10)&0.31(0.08)&          & 43(4) &            &1.24 &0.96(101)\\
\hline  
\noalign{\smallskip}
\textbf{PKS 0548-322}&   &     &     &     &          &          &          &       &            &     &         \\
50493003 & 20/02/99  & 5642&12439& 9355&1.53(0.07)&0.51(0.07)&2.88(0.27)& 75(5) & 153.6(3.2) & 2.02&1.21(67) \\
50493004 & 26/02/99  & ----& 2025& 1003&2.40(0.14)& -------- & -------- & 12(2) & ---------- & 1.76&1.31(29) \\
504930042& 07/04/99  & 5251&18943&10164&1.77(0.07)&0.45(0.06)&1.77(0.21)& 67(5) & 115.2(4.8) & 1.36&0.98(67) \\ 
\hline  
\noalign{\smallskip}
\textbf{1ES 1101-232}&   &     &     &     &          &          &          &       &            &     &         \\
50064017 & 04/01/97  & 6146&13868&10676&1.64(0.08)&0.33(0.07)&3.50(0.33)&120(7) & 241.6(3.2) &3.71 &1.08(184)\\
50726001 & 19/06/98  & ----&24817&10792&1.97(0.22)& -------- & -------- &145(4) & ---------- &2.55 &1.31(80) \\     
\hline 
\noalign{\smallskip}
\textbf{Mrk 180}&   &     &     &     &          &          &          &       &            &     &         \\
50064010 & 10/12/96  & 5165&18205& 7330&2.24(0.08)&0.28(0.08)&0.37(0.16)&36(4)  & 65.6(14.4) &0.51 &0.91(56) \\ 
\hline 
\noalign{\smallskip}        
\textbf{1ES 1218+304}&   &     &     &     &          &          &          &       &            &     &         \\
50863005 & 12/07/99  &10609&42693&20670&2.11(0.03)&0.38(0.03)&0.73(0.08)&98(3)  & 158.4(6.4) &1.48 &0.75(101)\\   
\hline 
\noalign{\smallskip} 
\textbf{1H 1426+428}&   &     &     &     &          &          &          &       &            &     &         \\
50493006 & 08/02/99  & ----&40657&20432&2.22(0.11)& -------- & -------- &103(16)& ---------- &2.04 &0.95(80) \\   
\hline 
\noalign{\smallskip}
\textbf{1ES 1553+113}&   &     &     &     &          &          &          &       &            &     &         \\
50064005 & 05/02/98  & 4421&10592& 4671&2.17(0.07)&0.63(0.08)&0.73(0.11)&115(8) &190.4(1.6)  &1.29 &1.22(67) \\
\hline 
\noalign{\smallskip}
\textbf{Mrk 501}&   &     &     &     &          &          &           &       &             &     &         \\
50377001$^*$ & 07/04/97  &12387&20571& 8936&1.68(0.01)&0.17(0.01)&8.7(1.3) &624(8) &1410(50)&21.5& -- \\ 
50377002$^*$ & 11/04/97  &12559&20391& 8719&1.64(0.01)&0.12(0.01)&31.6(9.6)&609(7) &1800(100)&23.9& -- \\   
50377003$^*$ & 16/04/97  & 9612&17125& 7347&1.41(0.01)&0.15(0.01)&101.6(23.7)&960(10)&6000(500)&52.4& -- \\   
50529001$^*$ & 28/04/98  &13552&21892& 9873&1.65(0.02)&0.15(0.02)&14.7(5.7)&474(8) &1200(100)&17.6& -- \\ 
50529002$^*$ & 29/04/98  &14661&21417& 9662&1.62(0.02)&0.17(0.02)&13.1(4.3) &543(8) &1400(100)&20.8& -- \\ 
50529003$^*$ & 01/05/98  &13164&19033& 8447&1.71(0.02)&0.24(0.02)&4.0(0.6) &477(8) & 940(30)&14.7& -- \\   
50666001$^*$ & 20/06/98  &98363&25861&11566&1.79(0.02)&0.21(0.02)&3.2(0.5) &175(4) & 320(10)  &4.95 & -- \\ 
50666002$^*$ & 29/06/98  &12682&47528& 7470&1.69(0.02)&0.23(0.02)&4.7(0.8) &323(6) & 660(30)&10.3& -- \\
50666003$^*$ & 16/07/98  &11228&15924& 6919&1.70(0.02)&0.33(0.02)&2.8(0.3) &320(7) & 600(20)&9.0 & -- \\ 
50666004$^*$ & 25/07/98  &25067&30937&14505&1.76(0.01)&0.28(0.01)&2.7(0.1) &337(5) & 610(10)  &9.3 & -- \\  
50944001$^*$ & 10/06/99  &10405&17510& ----&2.15(0.01)&0.24(0.01)&0.49(0.03) &186(2) & 315(4)  &3.02 & -- \\ 
\hline 
\noalign{\smallskip}
\textbf{1ES 1959+650}&   &     &     &     &          &          &          &       &            &     &         \\
50064002 & 04/05/97  & 2252&12391& 5393&2.02(0.18)& -------- & -------- &192(14)& ---------- &1.77 &0.90(56) \\  
51386001 & 23/09/01  &     &     &     &          &          &          &       &            &     &P.       \\  
513860011& 28/09/01  &25255&48037&22416&1.79(0.03)&0.43(0.02)&1.74(0.08)&464(9) &785.6(6.4)  &10.35&2.00(125)\\ 
\hline  
\noalign{\smallskip}
\textbf{PKS 2005-489}&   &     &     &     &          &          &          &       &            &     &         \\
50046001 & 29/09/96  & ----& 9911& 7674&2.02(0.19)& -------- & -------- &268(12)& -----      &5.94 &1.36(79) \\  
50503002 & 01/11/98  &20067&52467&23437&2.01(0.02)&0.17(0.02)&0.95(0.14)&821(12)&1313.6(19.2)&15.45&1.03(120)\\  
\hline  
\noalign{\smallskip}
\textbf{PKS 2155-304}&   &     &     &     &          &          &          &       &            &     &         \\
50016001 & 20/11/96  &35644&10686& ----&2.43(0.01)&0.24(0.01)&0.13(0.01)&504(4) &1240.0(28.8)&5.37 &1.22(127)\\
50160008 & 22/11/97  &22086&59497&28007&2.38(0.01)&0.37(0.01)&0.30(0.01)&780(8) &1569.0(27.2)&8.23 &1.75(125)\\
50880001 & 04/11/99  &45208&10392&49144&2.66(0.01)&0.20(0.01)& -------- &292(4) &1635.2(118.4)&2.48&1.39(125)\\  
\hline 
\noalign{\smallskip}
\textbf{PKS 2354-315}&   &     &     &     &          &          &          &       &            &     &         \\
50493007 & 21/06/98  &15055&40950&18034&1.77(0.03)&0.28(0.03)&2.57(0.22)&90(3)  & 160.0(1.6) &2.13 &1.33(101)\\   
\hline 
\noalign{\smallskip} 
\end{tabular}\\
$(^*)$ refers to Masaro et al. 2004
\end{table*}
\setcounter{table}{1}
\begin{table*}
\caption{\xmm~spectral analysis results.}
\begin{tabular}{|llllrrrrrrr|}
\hline
\noalign{\smallskip}
Obs ID&$Date$&Exps&Frame&$a$&$b$&$E_p$&$K$&$S_p$&$F_X$&$\chi^2_{r}$\\
\hline 
\noalign{\smallskip}
\textbf{1ES 0347-121}&   &     &         &          &          &          &       &            &    &         \\   
0094381101& 28/08/02 & 5362&M1-PW(Th)&          &          &          &       &       &    &F.       \\
\hline 
\noalign{\smallskip}
\textbf{PKS 0548-322}&   &     &         &          &          &          &       &            &    &         \\   
0111830201& 03/08/01 &47370&M1-PW(Md)&1.84(0.05)& -------- & -------- &95(1)  & -----      &3.11&0.95(124)\\ 
0205920501& 19/10/04 &40669&M2-PW(Th)&1.84(0.02)&0.14(0.03)&3.54(0.42)&126(1) & 224.0(1.6) &3.54&1.35(307)\\
\hline 
\noalign{\smallskip}
\textbf{1H 1100-230}&   &     &         &          &          &          &       &            &    &          \\      
0094380601& 29/05/01 & 2926&M1-PW(Md)&          &          &          &       &            &    &F.        \\  
0205920601& 08/06/04 &18267&M2-PW(Th)&2.04(0.02)&0.17(0.03)&0.76(0.06)&202(1) & 326.4(1.6) &4.10& 1.02(302)\\
\hline 
\noalign{\smallskip}
\textbf{Mrk 180}&   &     &         &          &          &          &       &            &    &          \\      
0094170101& 12/04/01 & 8294&M1-FW(Th)&2.27(0.03)& -------- & -------- &56(1)  & -----      &0.96& 0.87(82) \\
\hline 
\noalign{\smallskip}
\textbf{1ES 1218+304}&   &     &         &          &          &          &       &            &    &          \\       
0111840101& 11/06/01 &29309&M1-PW(Md)&2.19(0.03)&0.46(0.06)&0.63(0.08)&212(2) & 353.6(8.0) &2.68& 0.87(147)\\
\hline 
\noalign{\smallskip}  
\textbf{1H 1426+428}&   &     &         &          &          &          &       &            &    &          \\  
0111850201& 16/06/01 &60496&M1-PW(Md)&1.68(0.02)&0.12(0.02)&21.95(0.60)&82(1) & 212.8(6.4) &3.00& 0.99(343)\\
0165770101& 04/08/04 &65652&M1-PW(Th)&1.89(0.01)&0.31(0.02)&1.50(0.41)&102(1) & 166.4(1.6) &2.24& 0.96(317)\\
0165770201& 06/08/04 &68662&M1-PW(Md)&1.89(0.02)&0.38(0.04)&1.38(0.06)&106(1) & 172.8(1.6) &2.16& 0.98(202)\\  
0212090201& 24/01/05 &30163&M1-PW(Md)&1.90(0.02)&0.40(0.03)&1.33(0.04)&132(1) & 214.4(1.6) &2.64& 1.28(249)\\  
0310190101& 19/06/05 &46884&M1-PW(Md)&1.81(0.01)&0.23(0.02)&2.61(0.12)&189(1) & 331.2(1.6) &5.07& 1.23(341)\\
0310190201& 25/06/05 &44805&M1-PW(Md)&1.89(0.03)&0.29(0.05)&1.55(0.09)&148(1) & 241.6(1.6) &3.33& 1.07(184)\\
0310190501& 04/08/05 &45480&M1-PW(Md)&1.97(0.02)&0.34(0.04)&1.11(0.06)&143(1) & 228.8(1.6) &2.69& 1.01(222)\\
\hline 
\noalign{\smallskip}  
\textbf{1ES 1553+113}&   &     &         &          &          &          &       &            &    &          \\  
0094380801& 06/09/01 & 5801&M1-PW(Md)&2.09(0.03)& -------- & -------- &22(1)  & -----      &0.50& 1.30 (91)\\ 
\hline 
\noalign{\smallskip}  
\textbf{Mrk 501}&   &     &         &          &          &          &       &            &    &          \\  
0113060201& 13/07/02 & 8740&M2-FW(Tk)&2.05(0.05)&0.29(0.09)&0.81(0.21)&231(4) & 371.2(8.0) &4.10& 1.13(102)\\
0113060401& 14/07/02 &11880&M2-FW(Md)&2.18(0.02)& -------- & -------- &242(3) & -----      &4.73& 0.76(119)\\
\hline 
\noalign{\smallskip}  
\textbf{1ES 1959+650}&   &     &         &          &          &          &       &            &    &          \\  
0094380201& 23/01/02 &  306&M1-PW(Md)&1.72(0.02)& -------- & -------- &582(11)& -----      &20.3& 0.86(111)\\
0094383301& 16/01/03 & 2200&M1-PW(Md)&          &          &          &       &            &    &F.        \\
0094383501& 09/09/03 & 7085&M1-PW(Md)&          &          &          &       &            &    &F.        \\
\hline
\noalign{\smallskip}  
\textbf{PKS 2005-489}&   &     &         &          &          &          &       &            &    &          \\  
0205920401& 04/09/04 &12667&M2-PW(Th)&3.03(0.04)& -------- & -------- &19(1)  & -----      &0.12& 1.22 (63)\\
0304080301& 26/09/05 &21063&M1-PW(Th)&2.27(0.01)& -------- & -------- &123(1) & -----      &2.14& 0.96(201)\\   
0304080401& 28/09/05 &27662&M1-PW(Th)&2.28(0.01)& -------- & -------- &116(1) & -----      &1.96& 1.39(214)\\
\noalign{\smallskip} 
\hline
\end{tabular}
\end{table*}
\setcounter{table}{1}
\begin{table*}
\caption{{\it (continued)}~\xmm~spectral analysis results.}
\begin{tabular}{|llllrrrrrrr|}
\hline
\noalign{\smallskip}
Obs ID&$Date$&Exps&Frame&$a$&$b$&$E_p$&$K$&$S_p$&$F_X$&$\chi^2_{r}$\\
\hline 
\noalign{\smallskip} 
\textbf{PKS 2155-304}&   &     &         &          &          &          &       &            &    &          \\  
0124930101&30/05/00a &39301&M1-PW(Md)&2.49(0.02)&0.13(0.04)& -------- &393(1) &1806.4(592) &4.41&1.16(202)\\
0124930101&30/05/00b &39301&M1-PW(Md)&2.52(0.01)& -------- & -------- &406(3) & -----      &4.94& 0.88(158)\\ 
0124930201&31/05/00  &56894&M2-PW(Md)&2.55(0.01)& -------- & -------- &395(2) & -----      &4.60& 1.16(197)\\
0080940101&19/11/00a &57650&M2-PW(Th)&2.57(0.01)&0.21(0.02)& -------- &389(1) & -----      &3.66& 1.04(269)\\
0080940101&19/11/00b &57650&M2-PW(Th)&2.67(0.01)&0.18(0.03)& -------- &326(1) & -----      &2.76& 0.94(243)\\
0080940101&19/11/00c &57650&M2-PW(Th)&2.71(0.02)&0.23(0.04)& -------- &296(2) & -----      &2.27& 1.03(175)\\
0080940301&20/11/00a &58549&M2-PW(Th)&2.79(0.02)& -------- & -------- &253(2) & -----      &2.13& 0.93(137)\\
0080940301&20/11/00b &58549&M2-PW(Th)&2.71(0.02)&0.30(0.04)& -------- &269(2) & -----      &1.98& 1.14(169)\\ 
0124930301&30/11/01A &27749&M1-PW(Th)&2.48(0.01)&0.49(0.02)&0.32(0.02)&540(2) &1136.0(22.4)&4.56& 1.27(294)\\ 
0124930301&30/11/01B &38249&M1-PW(Md)&2.38(0.01)&0.48(0.01)&0.40(0.02)&779(2) &1483.2(16.0)&7.56& 1.27(357)\\
0124930301&30/11/01C &24650&M1-PW(Tk)&2.51(0.01)&0.47(0.02)&0.29(0.02)&601(2) &1316.8(35.2)&4.95& 0.97(301)\\ 
0124930501&24/05/02a &99165&M1-PW(Md)&2.38(0.01)&0.24(0.03)&0.17(0.04)&262(1) & 590.4(35.2)&3.10& 1.05(249)\\
0124930501&24/05/02b &99165&M1-PW(Md)&2.29(0.01)&0.30(0.02)&0.33(0.04)&270(1) & 508.8(12.8)&3.41& 1.18(286)\\
0124930501&24/05/02c &99165&M1-PW(Md)&2.04(0.01)&0.34(0.02)&0.86(0.04)&392(2) & 628.8(3.2) &6.75& 1.23(308)\\
0124930601&29/11/02Aa&57751&M1-PW(Md)&2.53(0.01)&0.44(0.03)&0.25(0.03)&192(1) & 441.6(19.2)&1.58& 1.00(219)\\ 
0124930601&29/11/02Ab&57751&M1-PW(Md)&2.48(0.02)&0.44(0.03)&0.29(0.04)&184(1) & 395.2(16.0)&1.61& 1.15(214)\\
0124930601&29/11/02B &55606&M1-PW(Md)&2.42(0.01)&0.44(0.02)&0.33(0.02)&263(1) & 529.6(11.2)&2.50& 1.14(298)\\
0158960101&23/11/03a &26862&M1-PW(Tk)&2.73(0.02)&0.16(0.05)& -------- &171(1) & -----      &1.38& 1.16(162)\\   
0158960101&23/11/03b &26862&M1-PW(Tk)&2.81(0.02)& -------- & -------- &152(1) & -----      &1.25& 0.86(130)\\  
0158960901&22/11/04a &28662&M1-PW(Th)&2.73(0.02)&0.40(0.04)&0.12(0.03)&190(1) & 656.0(68.8)&1.24& 1.06(177)\\
0158960901&22/11/04b &28662&M1-PW(Th)&2.68(0.02)&0.36(0.06)& -------- &168(1) & -----      &1.22& 1.07(144)\\
0158961001&23/11/04a &40163&M1-PW(Th)&2.61(0.03)&0.33(0.07)& -------- &241(2) &745.6(137.6)&1.96& 1.03(132)\\  
0158961001&23/11/04b &40163&M1-PW(Th)&2.63(0.03)&0.33(0.07)& -------- &236(2) & -----      &1.87& 0.97(121)\\  
0158961001&23/11/04c &40163&M1-PW(Th)&2.61(0.02)&0.36(0.05)&0.14(0.04)&255(2) & 739.2(86.4)&2.02& 1.02(165)\\
0158961101&12/05/05  &27838&M1-PW(Th)&2.54(0.02)& -------- & -------- &323(3) & -----      &3.81& 1.01(120)\\
0158961301&30/11/05a &60163&M1-PW(Md)&2.46(0.02)&0.21(0.03)& -------- &338(2) & -----      &3.67& 1.07(226)\\
0158961301&30/11/05b &60163&M1-PW(Md)&2.47(0.02)&0.26(0.04)& -------- &324(2) & 854.4(83.2)&3.34& 1.08(207)\\
0158961301&30/11/05c &60163&M1-PW(Md)&2.49(0.01)&0.22(0.02)& -------- &384(1) & -----      &3.97& 1.34(304)\\
0158961401&01/05/06  &64562&M1-PW(Md)&2.45(0.01)&0.11(0.02)& -------- &140(1) & -----      &1.68& 1.28(290)\\ 
\hline 
\noalign{\smallskip}  
\textbf{1H 2356-309}&   &     &         &          &          &          &       &            &    &          \\  
0304080601&15/07/05  &18082&M2-FW(Th)&2.05(0.03)&0.22(0.06)&0.78(0.17)&49(1)  & 8.0(0.2)   &9.35& 1.11(156)\\
\noalign{\smallskip} 
\hline
\end{tabular}
\end{table*}
\setcounter{table}{2}
\begin{table*}
\caption{\swf~spectral analysis results.}
\begin{tabular}{|llllccccccc|}
\hline
\noalign{\smallskip}
Obs ID&$Date$&Frame&Exps&$a$&$b$&$E_p$&$K$&$S_p$&$F_X$&$\chi^2_{r}$\\
\hline 
\noalign{\smallskip}
\textbf{1ES 0347-121}&   &    &      &          &          &          &       &            &     &         \\     
00030808001&03/10/06 & pc & 3187 &2.18(0.02)& -------- & -------- &102(2) & ----       &2.25 &0.76(24) \\ 
\hline 
\noalign{\smallskip}
\textbf{PKS 0548-322}&   &    &      &          &          &          &       &            &     &         \\     
00044002001&13/12/04 & pc & 9310 &1.68(0.03)&0.26(0.05)&4.13(0.47)&121(2) & 234.2(6.4) &3.82 &1.06(155)\\ 
00044002273&14/01/05 & pc & 1116 & -------- & -------- & -------- & ----- & -----      &5.23 & ------- \\ 
00044002005&14/01/05 & pc & 5704 &1.71(0.04)&0.34(0.08)&2.72(0.41)&107(2) & 198.4(6.4) &2.98 &1.07(87) \\ 
00044002008&10/03/05 & pc & 4248 &1.69(0.04)&0.34(0.09)&2.89(0.30)&109(3) & 206.4(6.4) &3.12 &1.14(65) \\ 
00035008001&01/04/05 & pc & 1256 & -------- & -------- & -------- & ----  & -----      &3.63 & ------- \\ 
00035008001&01/04/05 & wt & 1395 &1.85(0.04)& -------- & -------- &127(3) & -----      &3.57 &0.82(76) \\ 
00035008002&27/04/05 & pc & 5194 &1.69(0.04)&0.28(0.08)&3.64(0.61)&124(3) & 243.2(8.0) &3.79 &1.19(92) \\ 
00035008002&27/04/05 & wt & 1823 &1.81(0.04)& -------- & -------- &113(3) & -----      &4.11 &1.16(90) \\ 
00035008003&13/05/05 & pc & 3349 &1.69(0.04)& -------- & -------- &103(3) & -----      &3.25 &1.07(66) \\ 
00035008005&21/05/05 & pc & 9187 &1.68(0.03)&0.38(0.06)&2.66(0.17)&134(2) & 252.8(4.8) &3.75 &0.95(119)\\ 
00044002268&21/05/05 & pc &40031 &1.83(0.03)&0.22(0.05)&2.45(0.20)&150(2) & 259.2(3.2) &3.95 &1.02(289)\\ 
00035008006&24/05/05 & pc & 1344 & -------- & -------- & -------- & ----  & ----       &3.95 & ------- \\ 
00035008007&26/05/05 & wt &  817 &1.87(0.05)& -------- & -------- &125(4) & ----       &4.06 &1.05(39) \\ 
00035008008&29/05/05 & pc & 3852 &1.72(0.03)&0.41(0.06)&2.22(0.20)&126(2) & 225.6(4.8) &3.23 &1.31(111)\\ 
00044002274&24/06/05 & pc & 7910 &1.69(0.03)&0.31(0.06)&3.21(0.40)&147(3) & 281.6(6.4) &4.35 &1.26(133)\\ 
00066004010&11/01/06 & pc & 1723 & -------- & -------- & -------- & ----  & ----       &2.85 & ------- \\ 
00030836001&28/11/06 & pc & 4260 &1.67(0.05)&0.52(0.11)&2.10(0.24)& 87(3) &158.4(4.8)  &2.16 &1.26(51) \\ 
00044002034&13/03/07 & pc & 1169 & -------- & -------- & -------- & ----  & ----       &2.18 & ------- \\ 
\hline 
\noalign{\smallskip}
\textbf{1ES 1011+496}&   &    &      &          &          &          &       &            &     &         \\  
00035012002&19/06/05&pc&7966& 2.34(0.03) & 0.50(0.09) & 0.46(0.08) & 663(2) & 1210(4) & 6.67 & 1.14( 75) \\
00035012003&26/06/05&pc&9123& 2.13(0.03) & 0.33(0.09) & 0.64(0.12) & 736(2) & 1211(3) & 1.14 & 0.83( 80) \\ 
00035012003&26/06/05&wt& 806& -------- & -------- & -------- & ----  & ----       & 1.02 & -------- \\     
00035012004&20/12/05&pc&7644& 2.26(0.03) & 0.47(0.10) & -------- & 936(2) & -------- & 1.07 & 1.23( 76) \\ 
\hline 
\noalign{\smallskip}
\textbf{1H 1100-230}&   &    &      &          &          &          &       &            &     &         \\  
00035013001&30/06/05 & pc & 8521 &1.93(0.02)&0.40(0.05)&1.22(0.07)&230(3) & 369.6(4.8) &4.34 &0.92(194)\\ 
00035013002&13/07/05 & pc & 2295 &1.99(0.10)& -------- & -------- &229(6) & -----      &4.21 &0.94(50) \\ 
00035013003&04/11/05 & pc & 1163 &1.95(0.09)& -------- & -------- &203(10)& -----      &4.51 &0.81(19) \\ 
\hline 
\noalign{\smallskip}
\textbf{Mrk 180}&   &    &      &          &          &          &       &            &     &         \\ 
00035015001&16/04/06 & pc & 2932 &2.20(0.04)&0.58(0.12)&0.67(0.10)&116(4) & 193.6(6.4) &1.31 &1.09(44) \\ 
00035015002&18/04/06 & pc & 6373 &2.17(0.03)&0.40(0.07)&0.57(0.11)&142(3) & 182.4(4.8) &1.94 &1.14(113)\\ 
\hline 
\noalign{\smallskip}
\textbf{1ES 1218+304}&   &    &      &          &          &          &       &            &     &         \\ 
00035016002&30/10/05 & pc & 2013 &1.97(0.06)& -------- & -------- &121(5) & -----      &2.39 &1.18(30) \\ 
00035016001&31/10/05 & pc & 3701 &2.07(0.04)&0.39(0.11)&0.81(0.13)&111(3) & 179.2(4.8) &1.76 &1.11(56) \\ 
00030376001&08/03/06 & pc & 3082 &2.15(0.06)& -------- & -------- &62(3)  & -----      &0.80 &0.89(25) \\ 
00030376002&09/03/06 & pc & 3149 &2.25(0.06)& -------- & -------- &51(3)  & -----      &0.88 &0.96(23) \\ 
00030376003&18/05/06 & pc & 1670 & -------- & -------- & -------- & ----- & -----      &2.09 & ------- \\ 
00030376004&19/05/06 & pc & 1448 & -------- & -------- & -------- & ----- & -----      &1.87 & ------- \\ 
00030376005&20/05/06 & pc & 1330 & -------- & -------- & -------- & ----- & -----      &2.22 & ------- \\ 
00030376006&21/05/06 & pc & 2223 &1.89(0.05)&0.44(0.14)&1.32(0.16)&115(4) & 187.2(8.0  &2.22 &0.56(34) \\ 
\hline 
\noalign{\smallskip}  
\textbf{1H 1426+428}&   &    &      &          &          &          &       &            &     &         \\ 
00035020001&30/03/05 & pc & 1056 & -------- & -------- & -------- & ----- & -----      &0.55 & ------- \\ 
00035020001&30/03/05 & wt & 1879 &1.99(0.05)& -------- & -------- &64(2)  & -----      &1.64 &1.19(57) \\ 
00035020003&02/04/05 & wt & 3425 &2.03(0.02)& -------- & -------- &137(2) & -----      &3.52 &0.90(178)\\ 
00051000002&19/06/05 & pc &21375 &1.75(0.02)&0.31(0.03)&2.49(0.16)&213(2) & 382.4(4.8) &5.71 &0.92(276)\\ 
00051000003&25/06/05 & pc &22818 &1.89(0.02)&0.34(0.04)&1.47(0.08)&174(2) & 284.8(4.8) &3.76 &1.37(198)\\ 
00030375001&07/03/06 & pc & 2064 &1.76(0.05)&0.49(0.10)&1.77(0.18)&195(6) & 334.4(11.2)&4.37 &0.57(51) \\ 
00030375002&07/03/06 & pc &  925 & -------- & -------- & -------- & ----- & -----      &4.13 & ------- \\ 
00030375003&20/03/07 & pc & 2343 &1.86(0.03)& -------- & -------- &118(4) & -----      &2.56 &1.28(35) \\ 
\hline 
\noalign{\smallskip}  
\textbf{1ES 1553+113}&   &    &      &          &          &          &       &            &     &         \\ 
00035021001&20/04/05 & pc & 5167 &2.21(0.03)&0.36(0.07)&0.51(0.11)&157(3) & 268.8(8.0) &2.10 &1.29(95) \\ 
00035021002&06/10/05 & pc & 8528 &2.14(0.02)&0.24(0.04)&0.52(0.09)&435(5) & 728.0(11.2)&7.16 &1.04(221)\\ 
00035021002&06/10/05 & wt & 2236 &2.21(0.02)& -------- & -------- &432(5) & ---------- &7.29 &0.94(215)\\ 
00035021003&07/10/05 & pc & 9093 &2.11(0.02)&0.23(0.04)&0.57(0.09)&387(5) & 640.0(9.6) &6.66 &1.04(216)\\ 
00035021003&07/10/05 & wt & 1587 &2.21(0.02)& -------- & -------- &407(6) & ---------- &6.75 &0.92(175)\\ 
\noalign{\smallskip} 
\hline
\end{tabular}
\end{table*}
\setcounter{table}{2}
\begin{table*}
\caption{{\it (continued)}~\swf~spectral analysis results.}
\begin{tabular}{|llllccccccc|}
\hline
\noalign{\smallskip}
Obs ID&$Date$&Frame&Exps&$a$&$b$&$E_p$&$K$&$S_p$&$F_X$&$\chi^2_{r}$\\
\hline 
\noalign{\smallskip} 
\textbf{Mrk 501}&   &    &      &          &          &          &       &            &     &         \\   
00035023001&21/04/05 & pc &  544 & -------- & -------- & -------- & ----- & ------     &4.82 & ------- \\ 
00035023002&18/06/05 & pc & 1828 &1.89(0.05)&0.56(0.11)&1.25(0.10)&476(14)& 771.2(24.8)&8.25 &0.85(51) \\ 
00030793001&18/07/06 & pc & 4465 &2.02(0.03)&0.34(0.07)&0.94(0.09)&298(6) & 476.8(4.8) &5.29 &1.06(114)\\ 
00030793002&19/07/06 & pc &  993 &2.16(0.06)& -------- & -------- &412(17)& -----      &5.66 &1.19(28) \\ 
00030793003&19/07/06 & pc & 1024 &2.12(0.06)& -------- & -------- &316(13)& -----      &5.37 &1.25(30) \\ 
00030793004&20/07/06 & pc & 2954 &2.01(0.03)&0.36(0.08)&0.97(0.09)&400(9) & 640.0(14.4)&7.12 &1.02(88) \\ 
00030793005&21/07/06 & pc & 2278 &1.99(0.03)&0.50(0.09)&1.02(0.07)&410(10)& 656.0(16.0)&6.53 &0.97(79) \\ 
00030793006&23/03/07 & pc & 2027 &2.22(0.04)& -------- & -------- &239(7) & -----      &3.67 &0.71(47) \\ 
00030793007&30/03/07 & pc & 2281 &1.90(0.04)&0.42(0.10)&1.33(0.13)&229(7) &371.2(11.2) &4.49 &0.73(55) \\ 
00030793008&07/04/07 & pc & 2038 &1.91(0.05)& -------- & -------- &191(7) & -----      &4.32 &0.82(51) \\ 
\hline 
\noalign{\smallskip}  
\textbf{1ES 1959+650}&   &    &      &          &          &          &       &            &     &         \\   
00035025001&19/04/05 & wt & 4433 &2.09(0.01)&0.36(0.03)&0.76(0.04)&744(5) &1203.2(8.0) &11.68&0.95(342)\\ 
00035025002&19/05/06 & wt & 1279 &1.99(0.02)&0.41(0.05)&1.02(0.07)&638(8) &1020.8(14.4)&10.90&1.02(205)\\ 
00035025003&21/05/06 & wt & 1990 &1.93(0.02)&0.37(0.04)&1.25(0.06)&747(7) &1204.8(12.8)&14.51&0.92(266)\\ 
00035025004&23/05/06 & wt & 5365 &1.95(0.01)&0.23(0.02)&1.28(0.05)&964(5) &1552.0(9.6) &20.85&1.04(418)\\ 
00035025005&24/05/06 & wt & 2319 &1.85(0.01)&0.39(0.03)&1.58(0.05)&1087(8)&1801.6(14.4)&23.38&0.95(346)\\ 
00035025006&25/05/06 & wt & 4381 &1.92(0.01)&0.36(0.02)&1.30(0.03)&970(5) &1569.6(9.6) &19.29&1.04(414)\\ 
00035025007&26/05/06 & wt & 4384 &1.97(0.01)&0.31(0.02)&1.10(0.04)&962(5))&1540.8(8.0) &18.69&1.07(411)\\ 
00035025008&27/05/06 & wt & 4273 &2.03(0.01)&0.47(0.03)&0.92(0.03)&890(5) &1425.6(8.0) &13.66&1.23(364)\\ 
00035025009&28/05/06 & wt & 4399 &2.01(0.01)&0.46(0.03)&0.97(0.03)&872(5) &1395.2(8.0) &13.84&1.02(360)\\ 
00035025010&29/05/06 & wt & 3306 &1.72(0.03)&0.75(0.03)&1.52(0.04)&675(6) &1145.6(9.6) &12.41&1.08(273)\\ 
\hline
\noalign{\smallskip}  
\textbf{PKS 2005-489}&   &    &      &          &          &          &       &            &     &         \\ 
00035026001&31/03/05 & wt & 2214 &2.96(0.03)& -------- & -------- &170(3) & -----      &1.17 &1.28(116)\\ 
00035026002&05/04/05 & wt & 5858 &3.14(0.05)& -------- & -------- &109(2) & -----      &0.63 &0.80(115)\\ 
00035026003&06/04/05 & pc &15745 &3.02(0.02)& -------- & -------- &97(2)  & -----      &0.54 &1.09(129)\\ 
00035026003&06/04/05 & wt & 2511 &3.09(0.09)& -------- & -------- &83(2)  & -----      &0.50 &1.09(48) \\ 
\hline 
\noalign{\smallskip} 
\textbf{PKS 2155-304}&   &    &      &          &          &          &       &             &     &         \\   
00035027001&17/11/05 & pc &  934 & -------- & -------- & -------- & ----- & -----       &1.51 & ------  \\ 
00035027002&11/04/06 & pc & 2600 &2.31(0.04)& -------- & -------- &261(8) & -----       &3.59 &1.15(61) \\ 
00035027003&16/04/06 & pc & 5664 &2.35(0.03)&0.44(0.07)&0.40(0.07)&235(5) & 440.0(16.0) &2.45 &0.87(104)\\ 
00035027004&20/04/06 & pc & 2386 &2.21(0.12)& -------- & -------- &315(10)& -----       &4.09 &0.98(43) \\ 
00035027005&30/04/06 & pc & 8064 &2.46(0.03)& -------- & -------- &146(3) & -----       &1.63 &0.88(96) \\ 
00030795001&29/07/06 & wt & 4916 &2.56(0.01)&0.22(0.02)& -------- &811(5) & -----       &7.66 &1.12(304)\\ 
00030795002&01/08/06 & pc & 1239 &2.45(0.04)&0.65(0.12)&0.45(0.07)&653(18)&1248.0(48.0) &5.06 &1.33(60) \\ 
00030795003&01/08/06 & wt & 1843 &2.67(0.01)&0.20(0.04)& -------- &565(6) & -----       &4.75 &1.16(189)\\ 
00030795004&03/08/06 & wt & 1605 &2.53(0.01)&0.24(0.04)& -------- &687(7) & -----       &6.63 &1.19(199)\\ 
00030795005&05/08/06 & wt &  516 &2.80(0.03)& -------- & -------- &495(12)& -----       &3.21 &0.78(83) \\ 
00030795006&06/08/06 & pc &  648 & -------- & -------- & -------- & ----- & -----       &2.71 & ------  \\ 
00030795008&08/08/06 & pc &  679 & -------- & -------- & -------- & ----- & -----       &4.36 & ------  \\ 
00030795008&08/08/06 & wt &  438 &2.72(0.03)& -------- & -------- &490(12)& -----       &3.98 &0.88(80) \\ 
00030795009&10/08/06 & pc &  709 &2.50(0.05)& -------- & -------- &493(19)& ------      &4.75 &0.77(35) \\ 
00030795010&11/08/06 & pc &  919 & -------- & -------- & -------- & ----- & -----       &2.15 & ------  \\ 
00030795012&13/08/06 & pc & 1131 & -------- & -------- & -------- & ----- & -----       &1.65 & ------  \\ 
00030795013&14/08/06 & pc &  931 & -------- & -------- & -------- & ----- & -----       &2.44 & ------  \\ 
00030795014&15/08/06 & pc &  899 & -------- & -------- & -------- & ----- & -----       &1.55 & ------  \\ 
00030795015&16/08/06 & pc &  943 & -------- & -------- & -------- & ----- & ------      &1.09 & ------  \\ 
00030795016&17/08/06 & pc & 1006 & -------- & -------- & -------- & ----- & ------      &1.19 & ------  \\ 
00030795017&18/08/06 & pc & 1116 & -------- & -------- & -------- & ----- & ------      &1.89 & ------  \\ 
00030795024&26/08/06 & pc &  904 & -------- & -------- & -------- & ----- & ------      &0.74 & ------  \\ 
00030795025&27/08/06 & pc &  943 & -------- & -------- & -------- & ----- & ------      &1.20 & ------  \\ 
\hline 
\noalign{\smallskip}  
\textbf{1ES 2344+514}&   &    &      &          &          &          &       &             &     &        \\        
00035031001&19/04/05 & pc & 4665 &1.45(0.14)&1.06(0.24)& -------- &48(3)  & ------      &0.98 &1.09(35)\\ 
00035031002&19/05/05 & pc & 4183 & -------- & -------- & -------- & ----  & ------      &1.20 & ------ \\ 
00035031003&03/12/05 & pc &12204 &1.72(0.01)& -------- & -------- &41(2)  & ------      &1.05 &0.98(53)\\ 
\noalign{\smallskip} 
\hline
\end{tabular}
\end{table*}
\end{document}